\documentclass[%
 reprint,
 superscriptaddress,
 amsmath,amssymb,
 aps]{revtex4-2}

\usepackage{graphicx}
\usepackage{dcolumn}
\usepackage{xcolor}
\usepackage[caption=false, ]{subfig}
\usepackage{bm}
\usepackage{hyperref}
\usepackage[mathlines]{lineno}
\usepackage{float}

\begin{document}

\title{Exploring the spatial segmentation of housing markets from online listings}

\author{David Abella$^*$,$^\dagger$}\affiliation{Instituto de F\'{\i}sica Interdisciplinar y Sistemas Complejos IFISC (CSIC-UIB), 07122 Palma de Mallorca, Spain}

\author{Johann H. Mart\'inez$^\dagger$}\affiliation{Instituto de Matem\'atica Interdisciplinar, Departamento de An\'alisis Matem\'atico y Matem\'aticas Aplicadas, and GISC, Universidad Complutense, 28040 Madrid, Spain}

\author{Mattia Mazzoli}\affiliation{ISI Foundation, via Chisola 5, 10126 Turin, Italy}

\author{Thibault Le Corre}\affiliation{D\'epartement de G\'eographie, Universit\'e de Montr\'eal, Montr\'eal, Canada}

\author{Julien Migozzi}\affiliation{School of Geography and the Environment, University of Oxford, Oxford, United Kingdom}

\author{Eduard Alonso-Paul\'{\i}}\affiliation{Departament d'Economia de l'Empresa, Universitat de les Illes Balears, 07122 Palma de Mallorca, Spain}

\author{Rafel Cresp\'{\i}-Cladera}\affiliation{Departament d'Economia de l'Empresa, Universitat de les Illes Balears, 07122 Palma de Mallorca, Spain}

\author{Thomas Louail}\affiliation{UMR 8504 G\'eographie-cit\'es (CNRS - EHESS - Universit\'e Panth\'eon-Sorbonne, Universit\'e Paris Cit\'e), Campus Condorcet, 93322 Aubervilliers, France}\affiliation{UMR 5194 PACTE (CNRS - Sciences Po Grenoble - Universit\'e Grenoble Alpes), 38000 Grenoble, France}

\author{Jos\'e J. Ramasco}\affiliation{Instituto de F\'{\i}sica Interdisciplinar y Sistemas Complejos IFISC (CSIC-UIB), 07122 Palma de Mallorca, Spain}

\begin{abstract} 
The real estate market shows an inherent connection to space. Real estate agencies unevenly operate and specialize across space, price and type of properties, thereby segmenting the market into submarkets. We introduce here a methodology based on multipartite networks to detect the spatial segmentation emerging from data on housing online listings. Considering the spatial information of the listings, we build a bipartite network that connects agencies and spatial units. This bipartite network is projected into a network of spatial units, whose connections account for similarities in the agency ecosystem. We then apply clustering methods to this network to segment  markets into spatially-coherent regions, which are found to be robust across different clustering detection algorithms, discretization of space and spatial scales, and across countries with case studies in France and Spain.  This methodology addresses the long-standing issue of housing market segmentation, relevant in disciplines such as urban studies and spatial economics, and with implications for policymaking. 
\end{abstract}
\maketitle

\noindent
$^\dagger$Equal contribution.\\
$^*$Corresponding author: david@ifisc.uib-csic.es

\section{Introduction}
\label{sec:introduction}

The spatial dimension of housing markets is a crucial aspect for urban studies and planning. Understanding the spatial segmentation of the housing market into submarkets \cite{morawakage2022housing,bourassa2003housing} has important implications for real estate valuation and investment decisions, which together affect urban development and social equity \cite{bourassa2003housing}. Spatial segmentation is the product of many factors such as residential location and the proximity to amenities \cite{bourassa2003housing}, differences in housing stock \cite{keskin2017defining}, price levels \cite{goodman1998housing}, and consumer preferences \cite{leishman2013predictive}. 

The spatial division of the real estate market has been studied from different perspectives and with different methods in the literature. Some studies have examined the spatial segmentation of the urban housing market focusing on neighborhood correlations of housing prices \cite{palm1978spatial}, the spatial effects of urban public policies on housing values \cite{baumont2009spatial}, the neighborhood quality and accessibility effects on housing prices \cite{dubin1992spatial}, while others have determined if a specific property market is spatially segmented into submarkets, and whether accounting for the existence of submarkets improves the accuracy of price modeling \cite{keskin2017defining,usman2021priori}. This is especially important for hedonic pricing models that seek to incorporate spatial autocorrelation and heterogeneity \cite{usman2021priori,paez2009recent,bitter2007incorporating, case2004modeling}. Ref. \cite{hu2022NovelApproach} distinguishes two main approaches for spatial segmentation: using pre-defined geographical boundaries based on \textit{a priori} knowledge, such as local administrative boundaries or expert areas used by market stakeholders, or relying on clustering methods to infer patterns from the structure of the data. For the latter, popular statistical approaches to divide space into submarkets are principal component analysis and hierarchical clustering \cite{goodman1998housing,bourassa1999defining,bourassa2003housing}.

The digitization of the housing market \cite{raehousing2024} provides untapped research opportunities for data-driven studies of market segmentation. With property portals being nowadays the dominant way to create and access market information, online listings constitute a new type of data to study housing markets \cite{sawyer1999ict, boeing2017new, boulay2021moving}. Scholars studied the spatio-temporal distribution of housing prices \cite{yao2018mapping,adolfsen_segmentation_2022}, revealed the persistence of spatial inequalities in the housing information landscape \cite{boeing2020online}, predicted the social profile of neighborhoods \cite{delmelle2021language}, or detected the segmentation of the market from online search patterns \cite{rae2015online}.
Aside price, pictures or textual descriptions, a listing includes a critical piece of information: the identity of the marketing agency that has posted the listing on the portal. As such, listings constitute digital traces \cite{salganikbit2017} of the work performed by real estate agencies when acquiring, selling or marketing on property portals. It is therefore possible to reconstruct, for each agency, its own portfolio of listings, whose volume and location patterns result from and reflect the heterogeneous practices and market shares of real estate agencies. By informing on \textit{who sells where}, 
listings offer new ways to examine how real estate agencies unevenly operate and specialize across space, thereby segmenting the market into submarkets \cite{palm1976RealEstate}.

\begin{figure}[h!]
\includegraphics[width=\columnwidth]{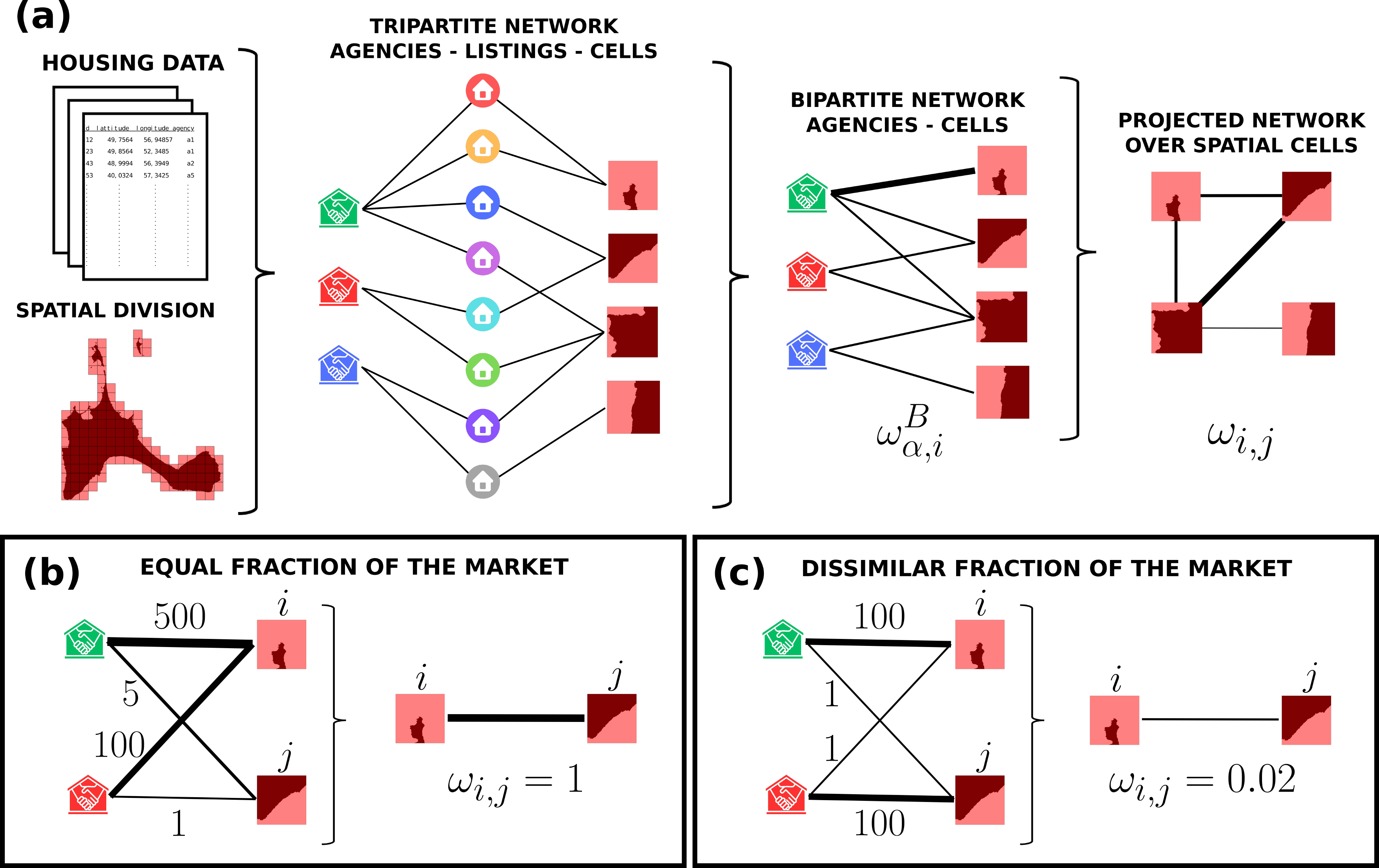}
\caption{\textbf{Bipartite network construction and projection.} (a) A tripartite network is constructed between real estate agencies, listings, and spatial units obtained from geolocalized housing data and the division of space in regular grid cells. In this network, each listing is connected to its real estate agency and the spatial cell where it is located. This simple tripartite network is contracted into a bipartite network linking agencies and cells, where the link weight $\omega^{B}_{\alpha,i}$ corresponds to the number of listings the agency $\alpha$ has in the spatial cell $i$. Finally, the network is then projected over the cells to form a weighed network of spatial units, where the weight $\omega_{i,j}$ of the link between cells $i$ and $j$ quantifies how much they are similar in the market -- $\omega_{i,j}$ is properly defined by Equation \eqref{eq:weight_eq}. Two simple examples of the projection process are shown below: with (b) equal and (c) complementary listings distributions for the agencies in the cells. }
\label{fig:network_construction}
\end{figure}

There is ample evidence underlining how real estate agencies influence market segmentation by determining housing prices, sorting homebuyers into different market channels, and specializing in certain types of neighborhoods and market segments \cite{palm1976RealEstate,palm1978spatial,keskin2017defining,bonneval2017agents,besbris2017investigating}. Furthermore, it has been shown that the definition of submarkets based on agencies is far superior to other segmentation techniques \cite{leishman2013PredictivePerformance}. 

This work introduces a new method to identify the housing market segmentation using geospatial data, complex network analysis techniques, and taking as a basis the local ecosystem of real estate agencies. We build a network structure based on two factors:  the \textit{presence} of an agency within a particular area, and the relative \textit{influence} of an agency in this area, determined by the agency's proportional share of all listings located in the area. Our methodology is applied to the residential property market in 3 Spanish provinces and 3 French urban areas, for which we have a rich, high resolution dataset sourced from property portals. We find that the market in those regions is divided into a hierarchy of subregions. We test the robustness of our results against different community detection algorithms, scales, and administrative boundaries in different countries.

\section{Materials and methods}
\label{sec:materials_and_methods}

\subsection{Data description}

For Spain, we analyze listings published on the portal \texttt{Idealista.com}~\cite{idealista}. The dataset covers a 2-year time period, from January 2017 to December 2018 and it comprises a comprehensive collection of online listings georeferenced with their (lat, long) coordinates in the Spanish provinces of Balearic Islands, Barcelona, and Madrid. These listings were posted by more than $50,000$ real estate agencies, each identified with its unique id. There are about one million listings for sales, and over $800,000$ for rentals. 

French listings were obtained from the portal \texttt{SeLoger.com}~\cite{SeLoger}. The dataset includes all listings posted in the country over a 6-month period from July to December 2019 - representing over 2 million sale listings. Geographical information is only available at the administrative and census levels, such as ZIP codes (``\textit{code postal}''), municipalities (``\textit{communes}''), and census tracts (``\textit{IRIS}''), the finest and basic scale for sub-municipal information in France. We focus on three major urban areas: Paris, Marseilles and Toulouse. 

For both datasets, we focus on houses and apartments, and do not consider farms or rural parcels.

\subsection{Building a network}

We begin by discretizing the space into spatial units (square grid cells, municipalities, districts, postal codes, census-tracts, etc).
This allows us to label each listing according to the spatial unit it falls into, along with the agency that posted this listing. By doing so, we build a tripartite relation between agencies, listings, and spatial units. Based on this structure, we can build a weighted bipartite network that connects agencies and spatial units, where the link weight $\omega^{B}_{\alpha,i}$ accounts for the number of listings posted by agency $\alpha$ that are located in the spatial unit $i$. The resulting network contains all the information about the spatial characteristics of the housing market. 

Bipartite networks can be projected to create networks with a single type of nodes \cite{newman2001structure,newman2001scientific,zhou2007bipartite}. In our case, we project it to build a new weighted network connecting spatial units (see Fig. \ref{fig:network_construction}(a) for schematic representation, taking as an example the discretization of space with square grid cells). Let us assume that we have $N$ spatial units and $N^a$ real estate agencies. The set of all agencies operating in the entire area is $\{ \alpha \}$, while the subset operating in the spatial unit $i$ is denoted by $\{ \alpha \}_i$. The fraction of listings in $i$ that belong to a certain agency $\alpha$ is
\begin{figure*}
    \begin{center}
    \includegraphics[width =0.8\textwidth]{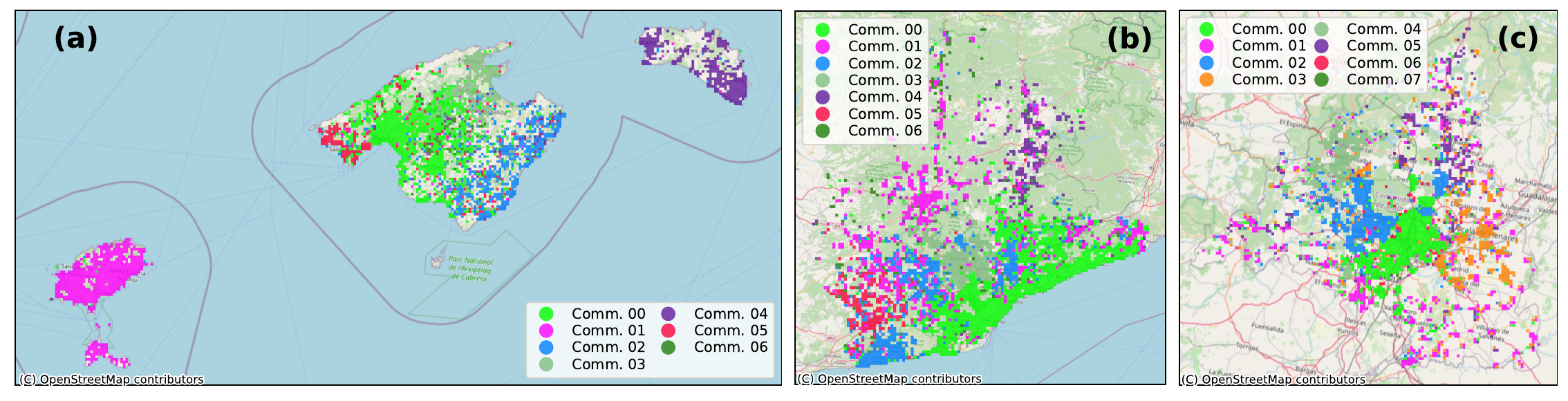}
    \caption{ \textbf{Market segmentation for $1 \; \textrm{km}$ square cells.} Communities from the projected network for the three Spanish provinces studied: Balearic Islands (a), Barcelona (b), and Madrid (c). The spatial cells are $1 \; \textrm{km}$ square cells. The communities shown are detected using the Louvain algorithm with a consensus clustering of $1000$ realizations. The underground map data is rendered from OpenStreetMap under ODbL.}
    \label{fig:cell_1000}
    \end{center}
\end{figure*}
\begin{equation}
    f_{\alpha, i} = \frac{\omega^{B}_{\alpha, i}}{\sum_{\gamma \in \{ \alpha \}_i} \omega^{B}_{\gamma, i}},
\end{equation}
where the index $\gamma$ runs over all the agencies operating in $i$. In the projected network, we define the influence weight between two spatial units $i$ and $j$ as
\begin{equation}
\omega_{i,j} = \frac{\sum_{\gamma \in \left\{ \alpha \right\}_{ij}} f_{\gamma,i} \, f_{\gamma,j}}{ \frac{1}{2} \left[ \sum_{\gamma \in \left\{ \alpha \right\}_{i}} f^2_{\gamma, i} + \sum_{\beta \in \left\{ \alpha \right\}_{j}} f^2_{\beta, j} \right]},
\label{eq:weight_eq}
\end{equation}
where $\{ \alpha \}_{ij} \equiv \{\alpha\}_i \cap \{\alpha\}_j$ is the subset of agencies operating in $i$ and $j$. The weight $\omega_{i,j} =1 $ if the agencies operating in $i$ and $j$ are the same, and cover an equal fraction of the market in both spatial units. If the market distribution is similar, but not equal, the weight will deviate from $1$. Reciprocally, if no common agency is found across the two spatial units, the weight is zero and there is no link between them. Fig. \ref{fig:network_construction}(b) and \ref{fig:network_construction}(c) show examples of the influence weights between two spatial units with equal distribution of the listings in (b), for which $\omega_{i,j} =1 $, and a complementary distribution in (c) with a value of $\omega_{i,j} = 0.02$. Note that our influence weight is related to the participation ratio introduced by Derrida \textit{et al.} in \cite{Derrida_1987}. 

The projected network is thus built with the spatial units as nodes, which are connected with links weighted according to Equation \ref{eq:weight_eq}. A group of spatial units strongly connected between them implies that they share a common ecosystem of agencies, that operate with a similar market share in these units.
Searching for clusters in this weighted spatial network should therefore inform us on the spatial segmentation of the housing market, the clusters corresponding to submarkets. In the network literature, such clusters are commonly referred to as communities, with numerous methods proposed to detect them \cite{fortunato2010community}. We use several classic community detection algorithms \cite{newman2004finding,infomap,Louvain,Louvain-Leiden,OSLOM} that account for network weights, including Louvain \cite{Louvain-Leiden}, Infomap \cite{infomap}, and OSLOM \cite{OSLOM}. These algorithms enable us to classify the spatial units into communities. Since these algorithms are stochastic, we perform several realizations of each method, and perform consensus clustering \cite{lancichinetti2012consensus} for higher stability.

\section{Results}
\label{sec:results}

\subsection{Segmenting the market according to agencies' operations}\label{sec:segmentation_cells}

We start by analyzing the spatial segmentation that arises from the data geolocated in the Balearic Islands, Barcelona, and Madrid using $1$ km-sided square cells. Fig. \ref{fig:cell_1000} presents the communities listed according to their size, from larger to smaller. Even though our methodology does not consider spatial proximity, we observe spatial segmentation in adjacent regions with few exceptions. For the Balearic Islands, we observe that spatial constraints, such as insular nature of the environment, affect the segmentation of the housing market: while the same submarket covers Minorca or Ibiza-Formentera, Majorca is divided into four different ones. It is noteworthy that the submarkets that emerge in all these three provinces are slightly larger than municipalities.

To study the robustness of identified submarkets in each of the three provinces, we run several community detection algorithms, and compare the communities obtained across realizations of different algorithms. We define as a network partition the classification of the cells in communities, $X = \{x_0, x_1, \cdots , x_{|X|-1} \}$, where each community $x_i$ is a set of cells. The partition $X$ has $|X|$ communities in this notation. Every cell must be in at least one community, but in some clustering methods a cell may belong to several.
In order to compare two partitions $X$ and $Y$, we compute a confusion matrix $C^{XY}$ in which each element is defined as
\begin{equation}
    C^{XY}_{ij} = | x_i \cap y_j | ,
\end{equation}
where $x_i$ and $y_j$ are communities in the partitions $X$ and $Y$, respectively, and $| . |$ stands for the cardinal (number of elements) of a set. An element $C^{XY}_{ij}$ can be zero if there is no overlap between the communities, and it can be large if the two communities coincide across the partitions. We reorder then the elements of the matrix  $C^{XY}$ to have the largest values in the pseudo-diagonal. 
\begin{figure}
    \begin{center}
    \includegraphics[width = \columnwidth]{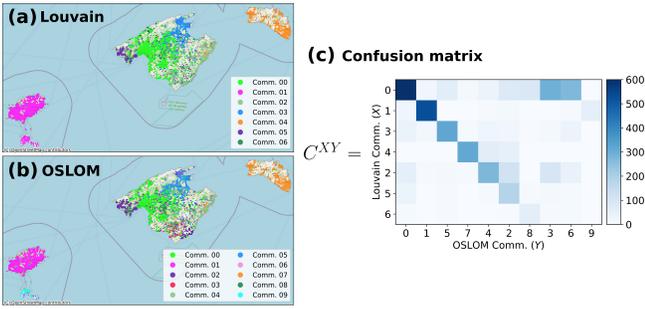}
    \caption{ \textbf{Agreement between different partitions.} Partition result of the community detection methods at the Balearic Islands using Louvain algorithm (a) and OSLOM method (b) $1 \textrm{km}$ square cells. The confusion matrix $C^{XY}$ of the two partitions (c), is ordered according to the maximum overlap. The underground map data is rendered from OpenStreetMap, under ODbL.}
    \label{fig:agreement_method}
    \end{center}
\end{figure}
Note that $C^{XY}$ is not necessarily a squared matrix because the number of communities in each partition may differ. This process is essentially the identification of the communities in one partition that correspond to the communities in the other. This is a statistical match, given that the cells of a community in $X$ may be distributed in several communities in $Y$. As shown in Fig. \ref{fig:agreement_method}, if the partitions between the two methods are similar, we must observe a strong pseudo-diagonal in the confusion matrix. The sum of the elements of this pseudo-diagonal is the number of cells clustered in the same way in the two partitions. To compute a measure of the agreement between two partitions, we use the fraction $H(X,Y)$ \cite{girvan2002community,hric2014community} defined as 
\begin{equation}
H(X,Y) = \sum_{i = 0}^{\textrm{min}(|X|,|Y|)-1} \frac{C^{XY}_{ii}}{N} , 
\end{equation}
where the matrix $C^{XY}$ is ordered to maximize the pseudo-diagonal, and $N$ is the total number of cells. $H(X,Y)$ is a metric commonly used in the literature to compute the accuracy between community detection algorithms \cite{danon2005comparing,duch2005community,li2008quantitative,darst2014improving,chen2015deep,saoud2016community,wang2017mitigation,fortunato2016community}, its value is bounded in the interval $(0,1]$, but it has the downside that $H(X,Y)$ depends on the size of the communities. To determine if the value of $H(X,Y)$ is significant, it is necessary to compare it with a randomized version of the partitions, $H(X_r,Y_r)$, in which the cells are reshuffled at random across the communities of each partition respecting the community sizes.

\begin{figure}
\begin{center}
\includegraphics[width = \columnwidth]{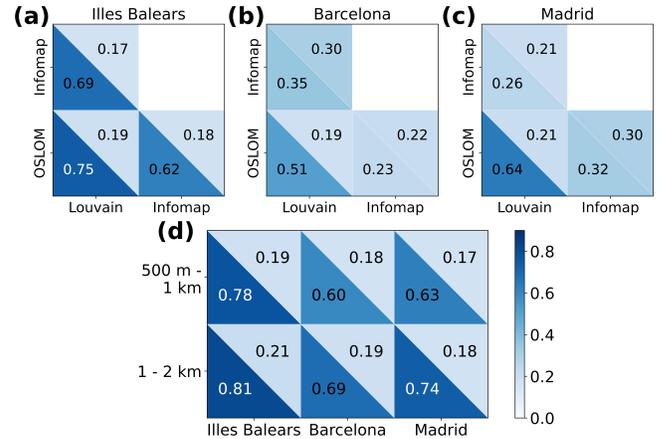}
\caption{ \textbf{Agreement across three methods and cell sizes.} Agreement across the different community detection methods for the network in Balearic Islands (a), in the province of Barcelona (b) and of Madrid (c). The metric used to compute the agreement between method partitions is $H(X,Y)$, shown in the lower triangles for each pair of methods, denoted by $X$ and $Y$. The upper triangles display the value $H(X_r,Y_r)$, being $X_r$ and $Y_r$ the partitions randomized (preserving the communities size). In (d), comparison of partitions obtained with the Louvain method for networks generated with different cell sizes: 500 m-sided vs  1 km-sided cells (top row), and 1 km-sided vs 2 km-sided cells (bottom row). }
\label{fig:agreement_scale_method}
\end{center}
\end{figure}

Figure \ref{fig:agreement_scale_method}(a-c) compares the three community detection algorithms (Louvain, OSLOM, and Infomap) used for different provinces. In all cases, the agreement between the communities detected from the real partition is higher than that of the randomized communities. The OSLOM-Louvain comparison exhibits the highest agreement, which is significant in all provinces. In the Balearic Islands, a robust and statistically significant agreement is evident among all methods. However, when examining Barcelona and Madrid, Infomap detects a large community probably due to the high density of the network, and this does not compare well with the other methods which detect more communities. In fact, the value of $H(X,Y)$ approaches the one of the randomized model. This issue is absent in the Balearic Islands, where the network has a stronger intrinsic spatial division into different islands.

\begin{figure}
\begin{center}
\includegraphics[width = \columnwidth]{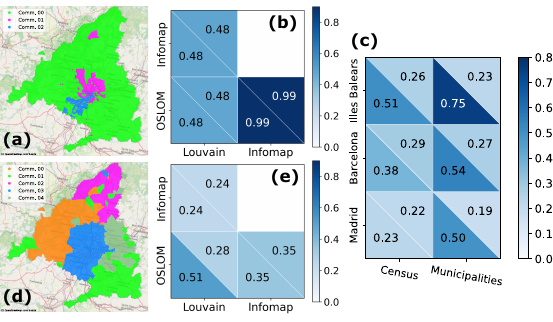}
\caption{ \textbf{Community detection from networks using administrative spatial units.} Communities detected using census areas (a) and municipalities (d) as spatial units to build the network in Madrid. The clustering method employed is the Louvain algorithm. The agreement across the different methods for the census (b) and municipalities (e). (f) shows the communities' agreement between $1$ km cells and administrative boundaries networks for all Spanish provinces. The agreement in (b)-(c)-(e) is computed using $H(X,Y)$ (lower triangles) compared with the value randomizing the communities (upper triangles).  The underground map data is rendered by OpenStreetMap, under ODbL.}
\label{fig:political_scales}
\end{center}
\end{figure}

So far, we have focused on the results for the networks built with $1 \; \textrm{km}$-sided square cells. It is, nevertheless, important to check whether the results may vary depending on the scale of the unit cells. We thus recalculate the networks taking as basis square cells of side $500 \; \textrm{m}$ and $2 \; \textrm{km}$ and compute the communities using the Louvain method with consensus clustering. 
The cells of the different scales have been delimited to keep spatial coherence: four $500 \;\rm{m}$ cells form one of the $1  \;\rm{km}$ cells used in the previous figures, and four $1 \; \rm{km}$ cells aggregate to form a $2 \; \rm{km}$ cell. 
This hierarchical structure allows us to compare communities at various levels because we can identify the cells across scales. For example, if a $2 \;\rm{ km}$ cell belongs to a community, then the four $1 \;\rm{ km}$ cells composing it share the same community label.
In parallel, we also run the community detection algorithm in the network composed of $1 \; \rm{ km}$ cells, and then we can use the confusion matrix and $H(X,Y)$ to compare the partitions at these two scales using $1 \;\rm{ km}$ cells. Note that the calculation of $H(X,Y)$ requires the same number of basic units in the two partitions. Figure \ref{fig:agreement_scale_method}(d) shows the results of this analysis, where we use $1$ km-sided cells as a reference for comparison with the other scales. In all cases, we notice a consistently high and statistically significant level of agreement. This demonstrates that our methodology generates communities that remain robust across the three spatial scales.

\subsection{Comparison with networks obtained from administrative boundaries}\label{sec:political spatial units}

In this section, we examine how incorporating administrative spatial boundaries to build networks impacts the detection of communities. In many cases, the geographical information for listings is only available at the level of existing administrative boundaries and statistical units, which are by design more heterogeneous than square cells.

We aggregate listings into administrative and statistical spatial units to determine if the emergent submarkets are stable and consistent when comparing with the ones observed with the networks built with square cells. In this case, we consider municipalities and census tracts as they are the most common administrative divisions applied to spatial statistics. 

Fig. \ref{fig:political_scales} shows the communities found in the province of Madrid. We observe clear differences between the results obtained using census tracts (Fig. \ref{fig:political_scales}(a)-(b) and using municipalities (Fig. \ref{fig:political_scales}(d)-(e). The results for census tracts are characterized by a large community that covers almost all the territory and the agreement between methods is not significant. In contrast, the results using municipalities have a good and significant OSLOM-Louvain agreement. Keeping Louvain as the reference method, we compare the partitions of the networks originated from $1  \, \textrm{km}$, census tracts, and municipalities in Fig. \ref{fig:political_scales}(c). The communities in the networks using cells and municipalities show significant agreement, while those based on census tracts show non-significant values in Barcelona and Madrid.

\begin{figure}
    \begin{center}
    \includegraphics[width = \columnwidth]{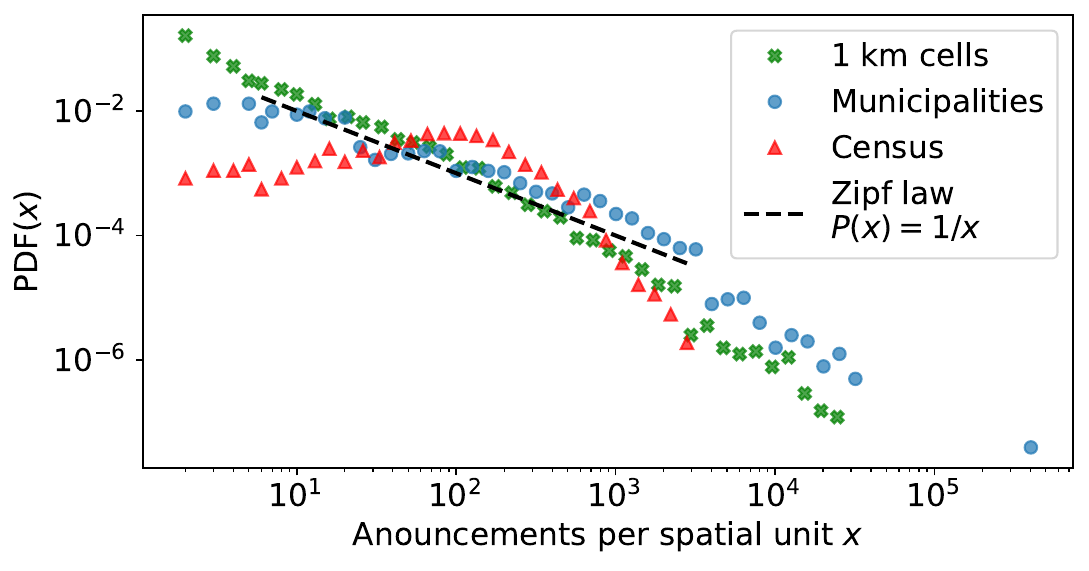}
    \caption{ \textbf{Distribution of listings for different spatial units.} Each spatial unit is shown by a different color and marker: green crosses (1 km-sided cells), blue circles (municipalities), and red triangles (census). The dashed black line shows the slope of a Zipf law distribution.}
    \label{fig:distribution_announcements}
    \end{center}    
    \end{figure}

While the distribution of listings per spatial unit in the other cases follows a heterogeneous distribution, well-described by a Zipf law, the one for census tracts follows a more homogeneous distribution (see Fig. \ref{fig:distribution_announcements}). This effect is a consequence of how the census tracts are built, forcing the population in each unit to be similar by a heterogeneous selection of the space included in each unit. This distribution is directly translated into the network weights and thus impacts the spatial segmentation method. 

\subsection{Recovering the submarkets from census level data}\label{sec:data_agreggation}

Multiple datasets, such as our French data, are avilable at census level. To maintain the broad applicability of our spatial segmentation methodology, we have devised a data aggregative method to recover the results obtained at the cell and municipality levels. This technique enables us to restore the Zipf law pattern using data gathered at the census level and to find similar segmentation results regardless of the basic spatial units.

\begin{figure}
\begin{center}
\includegraphics[width = \columnwidth]{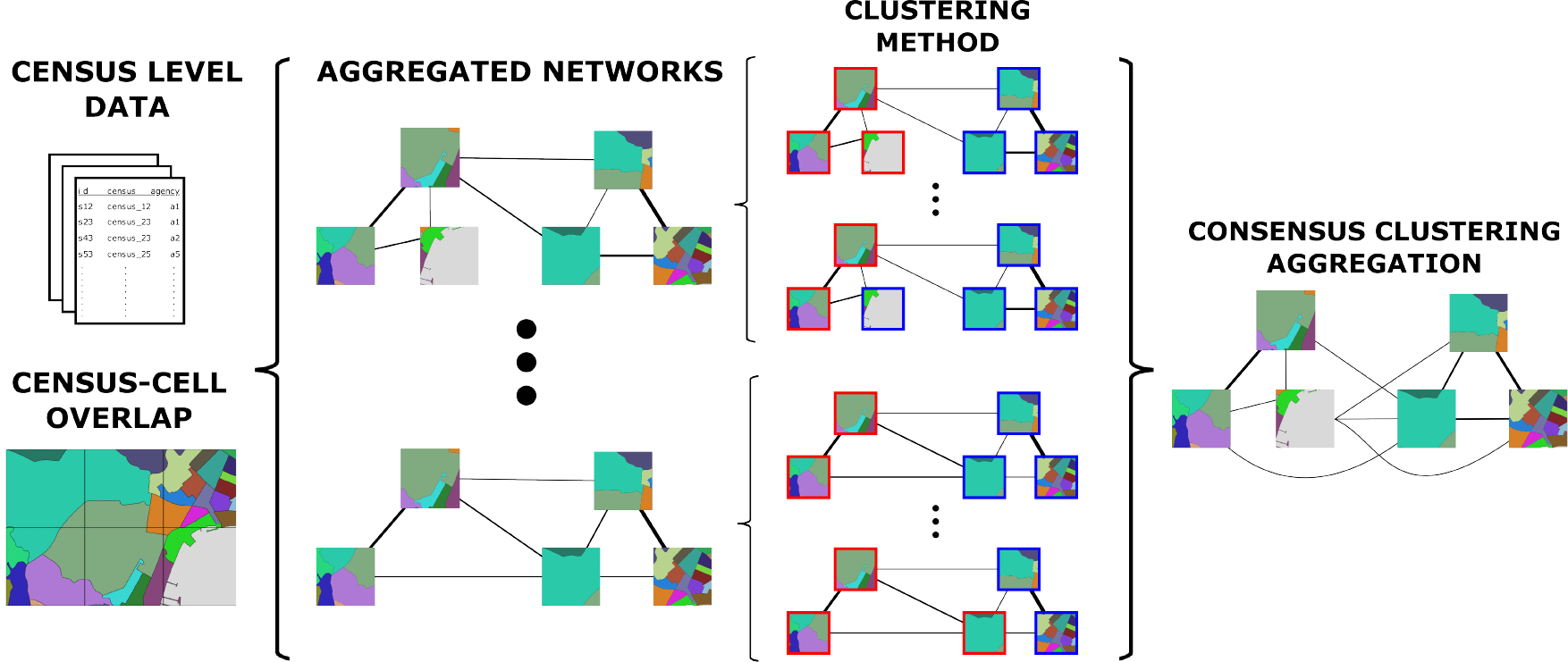}
\caption{ \textbf{Stochastic aggregative method using census level data.} From a census-level listing and the spatial division of the census in square cells, we generate an ensemble of networks. In this ensemble, each listing within a census tract is associated to a cell with a probability based on the overlapping area between the census and the cell. For each of these cell networks, we run a community detection algorithm multiple times. The next step involves combining the results from these partitioned networks through consensus clustering, resulting in an aggregated network.}
\label{fig:stochastic_construction}
\end{center}
\end{figure}

We start with listings at a census scale, such that each listing is associated to an agency and a census tract. The first step is to divide the space into square cells, as we did in Section \ref{sec:materials_and_methods}. The cells intersect with the census tracts. We then associate each listing to a cell with a probability proportional to the overlapping area between the listing census tract and the cell. This process is repeated for all the listings to reconstruct a tripartite network of agencies-listings-cells, from which we can follow the methodology explained to reach a cell-cell network and a segmentation in submarkets (communities). 
We observe that in the final networks the Zipf law distribution of listings per cell is recovered.

Since the assignation of listings to cells is stochastic, the projected network is different each time the process is repeated. To avoid uncertainty, we construct an ensemble of these networks. For each network, we run the community detection algorithm multiple times. Once our cells are labeled with a community, we perform consensus clustering to aggregate all partitions from all aggregated networks of our ensemble into a single consensus aggregated network. We represented this process in detail on Fig. \ref{fig:stochastic_construction}.

\begin{figure}
\begin{center}
\includegraphics[width =\columnwidth]{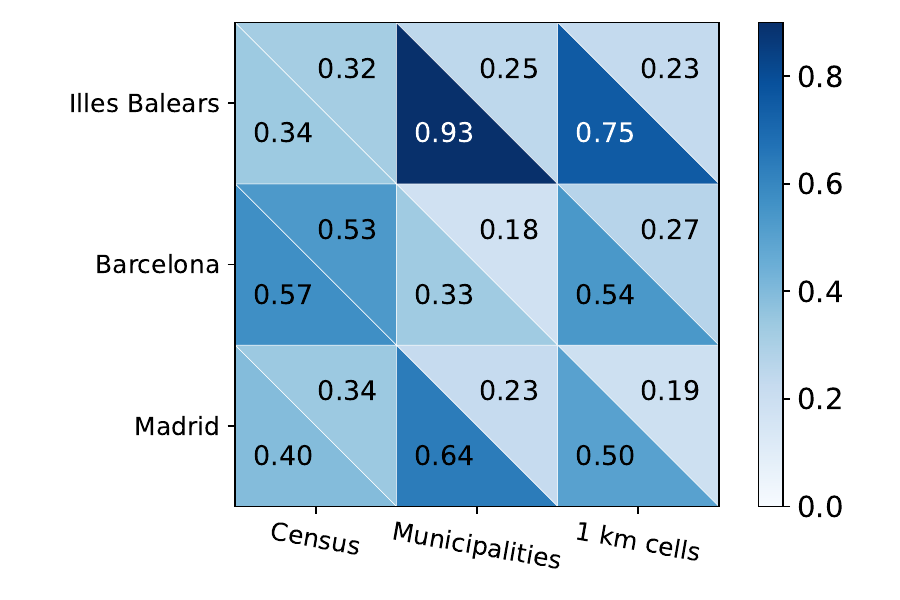}
\caption{ \textbf{Comparison between the communities from aggregated cells network and other spatial units.} Each column shows the agreement between the communities of the $1 \; \textrm{km}$ aggregated cells networks (from census data) and the networks obtained from the other spatial units: Census, Municipalities, and $1 \; \textrm{km}$ cells from the original latitude longitude coordinate data. Each row shows the results for each province: Balearic Islands, Barcelona and Madrid. The agreement is computed via the fraction of correctly detected cells $H(X,Y)$ (lower triangles) compared with the value randomizing the communities (upper triangles).}
\label{fig:network_sector_distribution}
\end{center}
\end{figure}

To verify the results of the aggregative method, we perform a comparison of the submarkets obtained out of different networks. Starting with our Spanish data, where the listings are geolocated using exact coordinates, we build networks at the level of $1 \, \rm{km}$ cells, census tracts and municipalities. We then apply the method to aggregate the census tracts to the cells. This gives us a fourth family of networks, which we call aggregated cells network. We then run community detection methods and compare them across the networks, taking as a basis the partition obtained from the network of aggregated cells (see Fig. \ref{fig:network_sector_distribution}). For all cases, the agreement exhibited by partitions of the aggregated cells network and the original cells or the municipalities is very high (and significant compared to the randomized communities). Therefore, by reconstructing the network with the aggregative method, we recover the original communities at the cell and municipality levels and avoid the issues caused by the natural spatial heterogeneity of census tracts.

\subsection{Comparison across countries}\label{sec:France}

In this section, we investigate whether the emergent spatial segmentation revealed by our method is a unique feature of the Spanish market, or can be understood as a more general phenomenon across geographical contexts. To this end, we use listing data for three major French urban areas, namely, Marseilles, Paris, and Toulouse. Since we do not have exact coordinates for the listings, which are only located at a census tract level, we have to employ the stochastic aggregative technique described in the previous section to obtain the networks at the cell level or to aggregate the data at the municipality (\textit{commune}) level (since the census tracts can be grouped within each commune). 

\begin{figure}
\begin{center}
\includegraphics[width =\columnwidth]{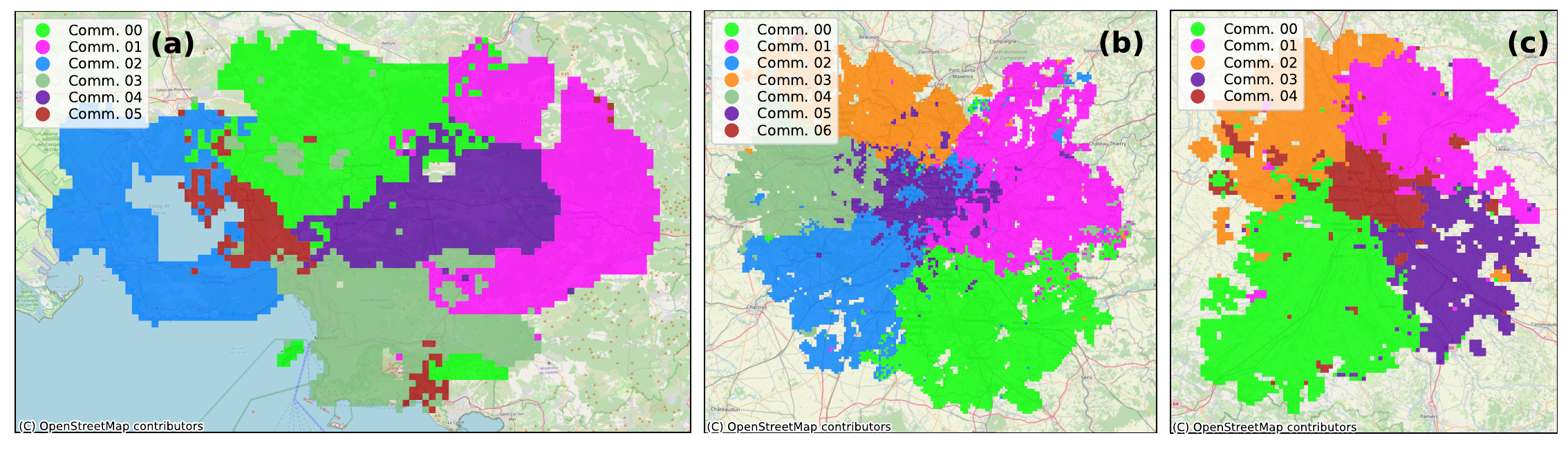}
\caption{ \textbf{Spatial segmentation for $1 \; \textrm{km}$ aggregated cells constructed from IRIS level data for France.} Communities detected at the stochastic projected network for the 3 French FUA studied: Marseilles-Aix en Provence (a), Paris (b), and Toulouse (c). The communities shown are detected using the Louvain algorithm with a consensus clustering of $200$ clustering method realizations for each of the $100$ stochastic networks generated in the IRIS to cell aggregative process. The underground map data is rendered by OpenStreetMap, under ODbL.}
\label{fig:france_comm}
\end{center}
\end{figure}

Communities emerge in these French urban areas at aggregated cell level as well (see Fig. \ref{fig:france_comm}). The communities are contiguous in space, similar to the ones observed in Spain, suggesting that listings (as a source of information on listed properties and agencies) allow us to study the spatial segmentation of the housing market through a data-driven, bottom-up method that foregrounds the practices of key market intermediaries.

\begin{figure}
    \begin{center}
    \includegraphics[width = \columnwidth]{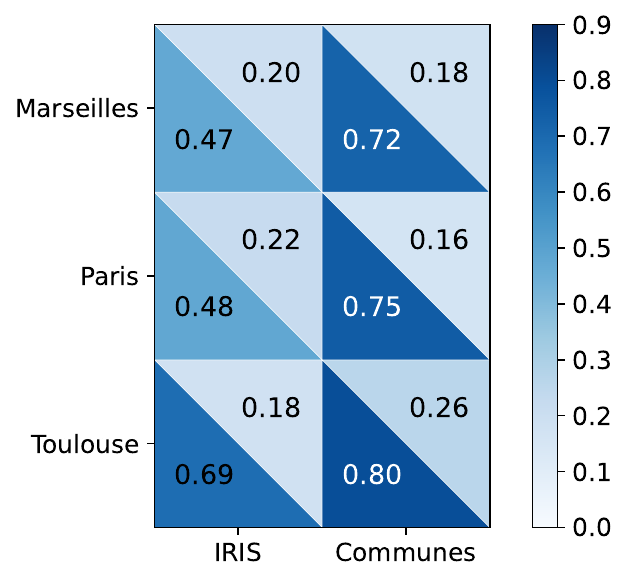}
    \caption{ \textbf{Comparison between the $1 \; \textrm{km}$ aggregated cells communities and the political units communities in France.} Each column shows the agreement between the $1 \; \textrm{km}$ aggregated cells and the French political spatial units: IRIS and Communes. Each row shows the results for each FUA: Marseilles-Aix en Provence, Paris, and Toulouse. The agreement is computed via the fraction of correctly detected cells (lower triangles) compared with the value randomizing the communities (upper triangles).}
    \label{fig:Compare_france}
    \end{center}
    \end{figure}

We repeat the exercise of comparing networks built from different spatial divisions. If France exhibits the same structures found in the Spanish dataset, we would expect the communities found from the aggregated cells and municipality networks to coincide, being the ones from the network of IRIS level very different. Fig. \ref{fig:Compare_france} displays the agreement between the communities using aggregated cells and administrative divisions (IRIS and communes). All values of the agreement are significant when compared with the randomized communities, but the largest agreement is found between aggregated cells and communes in all places, echoing results with the Spanish data. This indicates that our aggregative method is a general tool to compute a robust spatial segmentation of the housing market.

\section{Conclusions}
\label{sec:conclusions}

In this study, we present a new method for analyzing the spatial segmentation of housing markets through the activity of real estate agencies, using online listings to extract information on the location of both the property and the marketing agency. We apply this method to analyze comprehensive datasets of geolocated listings in two different countries: Spain and France. 

Our methodology is based on dividing space into spatial units, to construct a tripartite network between listings, real estate agencies, and spatial units. We project the network, taking into account the presence and influence of real estate agencies. To divide our projected networks, we use different classic community detection algorithms that account for network weights, such as Louvain, Infomap, and OSLOM. Our methodology generates a spatial segmentation into regions that happen to be spatially connected and larger than municipalities. This segmentation into submarkets remains robust across different community detection algorithms, scales, and administrative boundaries across different countries.

We discovered a limitation of our method when the spatial units exhibit a highly heterogeneous area distribution, and the Zipf law of the distribution of listings per spatial unit is not fulfilled, as in the case of census tracts. To overcome this limitation and extend our methodology to heterogeneous-level data, we developed a method to create an aggregated network via stochastic reconstruction and consensus clustering aggregation. This methodology exhibits good accuracy when compared with the communities from the original high-precision data.

To summarize, we have developed a new methodology that uses listings data to evaluate the spatial segmentation of housing markets into spatially-coherent submarkets. This methodology is generally applicable to different datasets of geolocated listings to infer the submarkets that emerge from the uneven presence and influence of real estate agencies across space. The market-based supra-municipal communities that emerge from the data are found to be robust. Future research should investigate how identifying the submarkets created by market intermediaries can inform policymaking and improve price modeling.

\section*{Author's contributions}
Conceptualization: DA, JHM, MM, TLC, JM, TL, JJR; Methodology and analysis: DA, JHM, MM, JJR; Data acquisition and curation: DA, JHM, MM, TLC, JM, EA-P, RC-C, TL, JJR; Writing (original draft preparation): DA, JJR; Writing (review and editing): DA, JHM, MM, TLC, JM, EA-P, RC-C, TL, JJR; Visualization: DA, JHM; Funding acquisition: RC-C, TL, JJR. All authors read and approved the final manuscript.

\section*{Acknowledgements}
DA, JHM, MM, EA-P, RC-C and JJR acknowledge funding from the CAIB (Government of the Balearic Islands) through the project NouLloguer (PRD2018/43). DA and JJR received partial funding from the Agencia Estatal de Investigaci\'on (AEI, MCI, Spain) MCIN/AEI/10.13039/501100011033 and Fondo Europeo de Desarrollo Regional (FEDER,UE) under project APASOS (PID2021-122256NB-C22) and the Maria de Maeztu Program for units of Excellence in R\&D, grant CEX2021-001164-M. JM, TL and TLC thank the \textit{Groupe SeLoger} for their precious collaboration and for making the data available through a partnership (Ref. CNRS N. 238072). The \textit{Groupe SeLoger} cannot be held responsible for the completeness, reliability and veracity of the results of this study.

\section*{Availability of data and materials}
The projected networks at the spatial resolution of $1$ km cell, census-tract, and municipality are available at Zenodo \cite{zenodo-2024} and Github \cite{Abella-github-2024}. These links also include the code and additional code to perform the stochastic aggregative method from generic census data.

\end{document}